\documentclass[aps,pra,twocolumn,groupedaddress,amsmath,amssymb,byrevtex]{revtex4-1}
\usepackage{graphicx}    
\usepackage{epstopdf}

\begin{document}

\title{Precise measurements on a quantum phase transition in antiferromagnetic spinor Bose-Einstein condensates}
\author{A. Vinit and C. Raman}
\email {chandra.raman@physics.gatech.edu}
\affiliation{School of Physics, Georgia Institute of Technology, Atlanta, Georgia 30332, USA}
\date{\today}

\begin{abstract}
We have experimentally investigated the quench dynamics of antiferromagnetic spinor Bose-Einstein condensates in the vicinity of a zero temperature quantum phase transition at zero quadratic Zeeman shift $q$.  The rate of instability shows good agreement with predictions based upon solutions to the Bogoliubov de-Gennes equations.  A key feature of this work was removal of magnetic field inhomogeneities, resulting in a steep change in behavior near the transition point.  The quadratic Zeeman shift at the transition point was resolved to 250 milliHertz uncertainty, equivalent to an energy resolution of $k_B \times 12$ picoKelvin.  To our knowledge, this is the first demonstration of sub-Hz precision measurement of a phase transition in quantum gases.  Our results point to the use of dynamics, rather than equilibrium studies for high precision measurements of phase transitions in quantum gases.  

\end{abstract}

\maketitle

Phase transitions are singular points in the behavior of many-body systems.  System properties change extremely rapidly in their vicnity, and in the thermodynamic limit the transition point becomes a singularity.  In the real world, the mathematical singularity is hidden by finite size effects and heterogeneity.  For example, in the domain of superfluids, the spectacular lambda point of liquid helium is smeared out by the earth's gravity, requiring that precision comparisons of experiment and theory  be performed in space \cite{lipa2003}.  For quantum gases of ultracold atoms this problem is even more severe, since density variations are intrinsic to the system \cite{pita03book}.  For example, in a Bose-Einstein condensate (BEC) the particle density varies by 100\% from the center to the edge of the Thomas-Fermi volume.   A phase transition that is controlled by the chemical potential will therefore be smeared out, typically by $k_B \times 100-300$ nanoKelvin.  

\begin{figure} [htbp]
\includegraphics[width= \columnwidth]{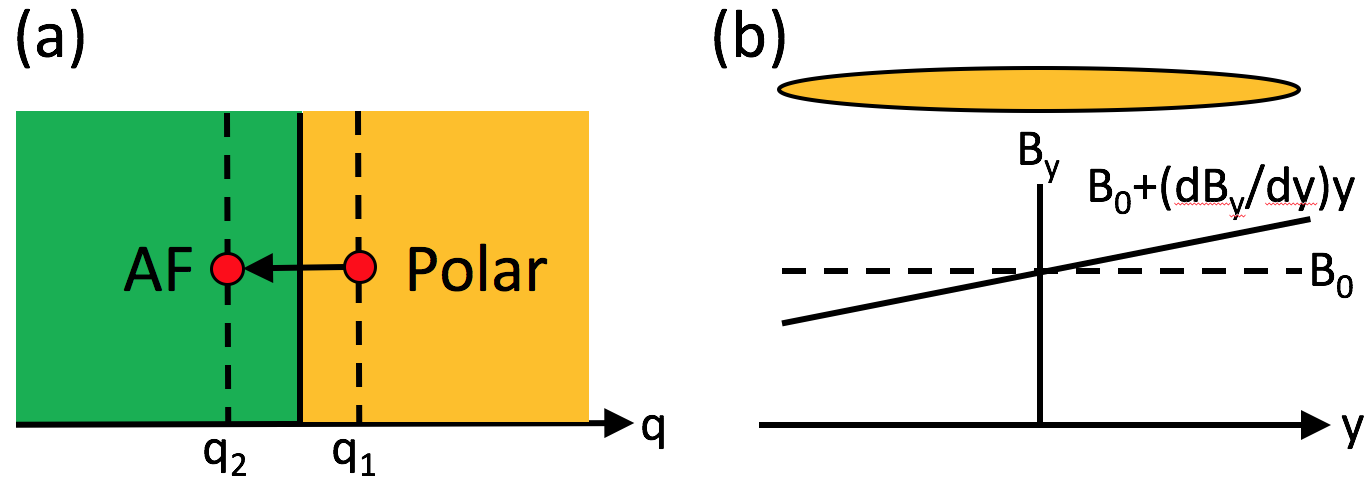}
\caption{(Color Online). Probing quench dynamics in an antiferromagnetic spinor BEC.  a) Instantaneous quench of quadratic Zeeman shift from $q_1 > 0$ to $q_2 < 0 $ through the quantum phase transition at $q=0$ is performed using an AC (microwave) magnetic field.  (b) DC magnetic field, kept constant during the experiment, consists of an applied bias field $\vec{B}$ that slowly varies in space, shown for fields applied along the long axis of the cigar-shaped condensate.}
\label{fig:diagram}
\end{figure}

Measuring the detailed behavior near a phase transition is important for revealing critical phenomena and universality, both of which are actively sought with cold atom systems \cite{bloch2012}. While local measurements on optical lattices in 2 dimensions \cite{bakr2010,weitenberg2011} have made great strides in alleviating the inhomogeneity problem, they are specialized geometries, and do not readily lend themselves to bulk matter in 3D.  In this paper we report precise measurements  of a quantum phase transition in a {\em bulk} spinor Bose-Einstein condensate that is controlled by a single energy scale, the quadratic Zeeman shift.  Spinor BECs are a rich arena for exploration due to the interplay of magnetic fields and magnetic interactions \cite{sten98spin,chan04,sadler06,black07,liu09,klempt2009,kronjager2010,zhao2015,seo2015}.  The transition we examine is between polar and anti-ferromagnetic spin states in a spin-1 $^{23}$Na BEC (see Figure \ref{fig:diagram}).  It is not smeared out by density inhomogeneities, as the critical point does not depend upon density at all.  Rather, the energy difference between the two competing states, which in turn is controlled by external fields through the quadratic Zeeman effect, sets the phase boundary of this first order transition \cite{bookjans2011,ueda2012review,stamper-kurn2012,phuc2013}.  Our measurements determine the location of the transition point with an unprecedented frequency uncertainty of $130$ mHz  and $ 220$ mHz due to statistical and systematic effects, respectively.  The combined error is equivalent to $k_B \times 12$ picoKelvin.

In this work we expand upon earlier observations \cite{bookjans2011,vinit2013} to make precise measurements of this phase transition.  A key factor enabling the enhanced precision reported here is the application of magnetic fields parallel to, rather than perpendicular to the long axis of the cigar-shaped BEC.  This has afforded us a tool to control and to explore the role played by magnetic field gradients in the quench dynamics.  Our data provide a powerful argument that these gradients were responsible for smearing out the phase boundary observed in earlier work \cite{bookjans2011}.  We argue that this arises from a decoherence mechanism inhibiting the production of spin pairs that tends to slow down the instability.  By removing the field gradient, we have measured an instability rate that is in good agreement with Bogoliubov theory, thus resolving discrepancies noted earlier \cite{bookjans2011}.  An important theme of our work is the use of {\em dynamics} to probe the phase transition boundaries, rather than attempting to reach the ground state through thermal equilibration, as other studies have done  \cite{sten98spin,liu09}.  

Our starting point is the spin-dependent mean-field Hamiltonian for spin $F = 1$ Bose-Einstein condensates in the low-energy spin sector, as written in Reference \cite{bookjans2011}, with an additional, linear Zeeman term (see \cite{stamper-kurn2012}):
\begin{equation}  H_{sp} =  \frac{c_2}{2} n({\bf r}) \langle \hat{\bf F}\rangle ^2 + p ({\bf r}) \langle \hat{F}_z \rangle + q \langle \hat{F}_z^2 \rangle  \label{eq:one} \end{equation}
For the low values of $p$ we are considering, the linear Zeeman term does not influence the overall density profile.  ${\bf \hat{F}},\hat{F}_z$ are the vector spin-1 operator and its $z$-projection, respectively and $n$ is the particle density.   For sodium atoms, the spin-dependent interaction coefficient $c_2= \frac{4 \pi \hbar^2}{3 M}(a_2-a_0) = + 1.6 \times 10^{-52} \rm{J/m^3} > 0 $ \cite{black07}, and hence the system is antiferromagnetic.  Here $M$ is the atomic mass, and $a_{2,0}$ are the triplet and singlet scattering lengths, respectively.  In our experiment, we control the linear Zeeman term through the spatial gradient of the magnetic field (see Figure \ref{fig:diagram}b), i.e., $p({\bf r}) = \mu B_0 + \mu \nabla B \cdot {\bf r}$ and the quadratic Zeeman shift $q = \tilde{q} B_0^2$.  $B_0$ is the magnetic field at the trap center, $\tilde{q} = 276 $ Hz/Gauss$^2$ is the coefficient of the quadratic Zeeman shift for sodium atoms, and the magnetic moment $\mu$ is $\frac{1}{2} \times $ the Bohr magneton $\mu_B$ \cite{ueda2012review}.  

For a perfectly homogeneous magnetic field, we may apply a gauge transformation to the Hamiltonian to set $p=0$ \cite{stamper-kurn2012}.  In this case, for an antiferromagnetic spinor BEC prepared in an initial state with zero net magnetization, as in our experiments, the ground state for $q > 0$ is a polar condensate consisting of a single component--the $m_F = 0$ spin projection that minimizes $\langle \hat{F}_z ^2 \rangle$.  For $q<0$ the ground state maximizes the same quantity through a superposition of two components $m_F = \pm 1$, a so-called antiferromagnetic phase \cite{kawaguchi2012}.  The symmetry properties of the ground state therefore change discontinuously at $q = 0$, defining a zero temperature quantum phase transition \cite{kawaguchi2012,bookjans2011}.  As in our earlier works, we used the AC Zeeman effect through a microwave magnetic field to vary $q$  \cite{bookjans2011,vinit2013}.

Optically confined, cigar-shaped Bose-Einstein condensates in the $m_F = 0$ state were prepared in a static magnetic field aligned with one of the coordinate axes $i = x,y,z$ depicted in Figure \ref{fig:diagram}.  The protocol is described in our earlier work \cite{bookjans2011}.  Axial Thomas-Fermi radius and trap frequency were measured to be $R_y = 355 \mu$m and $\omega_y = 2 \pi \times 7$ Hz, respectively.  From these we determined the peak spin-dependent interaction energy, $c_2 n_0 = h \times 110$ Hz, accurate to about 10\%.  From the known trap aspect ratio of 67, we estimated the radial Thomas-Fermi radius to be $R_\perp = $ 5 $\mu$m, which was small enough such that only axial spin domains could form.  The measured temperature was 400 nK, close to the chemical potential of 340 nK.  

We rapidly switched $q$ from $q_1 > 0$ to a final value $q_2< 0$ at $t=0$.  Following a variable hold time, we switched off the trap and used time-of-flight Stern-Gerlach (TOF-SG) observations to record the one-dimensional spatiotemporal pattern formation in each of the 3 spin components, $n_i(y); i = 0,\pm 1$, with a resolution of 10 $\mu$m.  For the current work we focus on the total population in each of the spin states, $N_i = \int n_i(y) dy$, as well as the population fractions $f_i = N_i/\sum_j N_j$.  
\begin{figure} [htbp]
\includegraphics[width= \columnwidth]{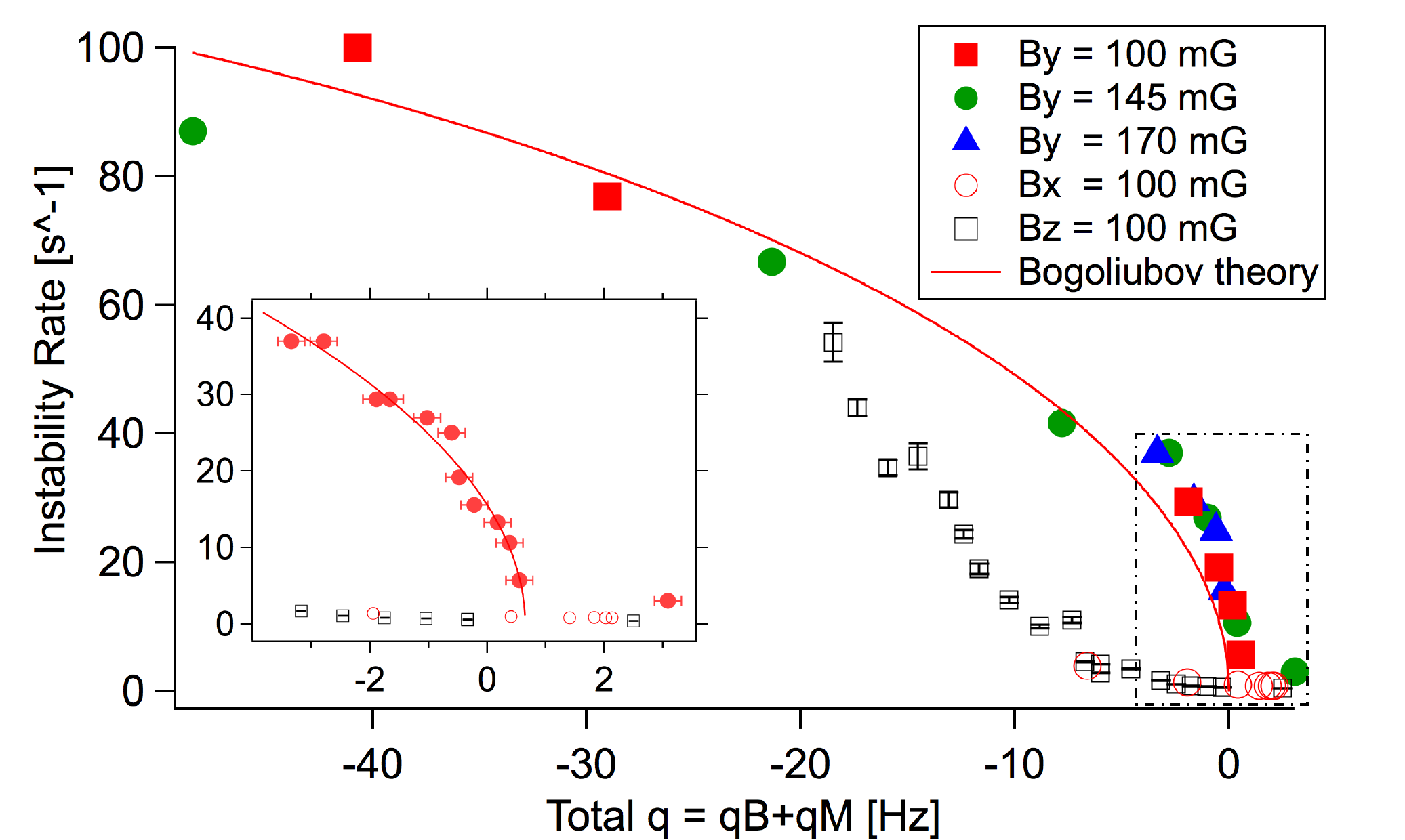}
\caption{(Color Online).  Probing the phase transition with $\sim100$ milliHertz sensitivity.  Solid symbols are data taken for different static magnetic fields $B_y$, i.e., along the cigar.  Clear symbols were taken in the two transverse directions $B_x,B_z$.  Solid line is a fit of all $B_y$ data to the rate given by Bogoliubov theory for a homogeneous BEC, as described in the text.  Inset is an expanded view of the data outlined in the dashed  box, with solid circles for all $B_y$ data.  Solid line is a separate fit allowing for an offset $q0$, measured to be $+0.65 \pm 0.13$ Hz.  Vertical error bars are statistical, while horizontal error bars are uncertainty in $q$-calibration of $0.22$ Hz.  $B_z$-data is reprinted from \protect{\cite{bookjans2011}. }}
\label{fig:Instability_Rate}
\end{figure}

To control magnetic bias fields during the previously described experimental sequence, we used three pairs of orthogonal Helmholtz coils wrapped around the vacuum chamber.  For a bias along $\hat{x}_j, j = 1,2,3$, we eliminated fields transverse to the $x_j$-axis with two pairs of coils, after which we applied a constant current along the $x_j$-axis.  The actual magnetic field magnitude $|B|$ at the location of the atoms was determined via microwave spectroscopy of the $F = 1, m_F = 0 \rightarrow F=2,m_F = \pm 1$ transitions through their Zeeman frequency shift of $g_F \mu_B |B|/h = 700$ kHz/Gauss relative to the ``clock'' transition $F = 1, m_F = 0 \rightarrow F=2,m_F = 0$ \cite{kase89}.  In addition to these bias coils, we used a single anti-Helmholtz coil pair aligned with the $y$-axis to generate magnetic field gradients of up to 160 mG/cm.  Background field gradients along the $y$-direction were observed to be in the neighborhood of 80 mG/cm.  Our uncertainty in magnetic field calibration  was $4$ mG.

In order to cancel the field gradient, we adjusted the current through these coils to maximize the lifetime, for $q > 0$, of the $m_F = 0$ component against spin relaxation into $\pm 1$ pairs, observing that the latter spin states were completely miscible under these conditions \cite{sten98spin}.  The finite current resolution of our power supply controls limited the resolution to $\pm 0.3$ mG/cm.  With this resolution the residual linear Zeeman energy $p$ after cancellation of stray magnetic field gradients was no more than $\pm 7$ \% of the spin-dependent interaction energy $U$. 

Figure \ref{fig:Instability_Rate} shows our main experimental observation.  It is the extreme sensitivity of the instability rate to quadratic Zeeman shift near the phase boundary, allowing for a precise determination of the latter.  Similar to our earlier work \cite{bookjans2011,vinit2013}, we measured the fractional population in $m_F = \pm 1$, i.e., $f_{\pm 1}=f_1+f_{-1}$, which was observed to grow with time.  The instability rate was defined to be $\Gamma_{1/2} = 1/T_{1/2}$, where $T_{1/2}$ was the time at which $f_{\pm 1}$ had increased to 1/2.  The extreme sensitivity was only observed for magnetic fields aligned parallel to the long axis of the cigar-shaped BEC (hereafter, $B_y$), where we could carefully null stray magnetic field gradients.  By contrast, we measured a stark difference for fields aligned along $B_{x,z}$.  In this case, according to our earlier observations (data reproduced from \cite{bookjans2011} in Figure \ref{fig:Instability_Rate} as open squares), a significant discrepancy was noticed between the experimentally observed instability rates and those predicted by Bogoliubov theory, particularly for data taken near the transition point.  The experimental data suggested a smooth turn-on of the instability rate rather than a sharp transition point.  In the current work we have reproduced this difference for magnetic fields $B_x$ (open circles), which agrees with the previous $B_z$ data.  With the newer $B_y$ data we observe a much closer agreement with the theoretical prediction for a homogeneous system, $\Gamma(q) = \sqrt{|q|(q+2 c_2 n_0)}$, the solid line in the Figure.  Here $n_0$ is the peak density of the $m_F=0$ cloud.  

Focusing now only on $B_y$ data with the gradients cancelled in Figure \ref{fig:Instability_Rate}, we applied 3 different magnetic fields $B_y = 100, 145$ and $170$ mG, and adjusted the microwave power accordingly to cover the same range of total quadratic Zeeman shift.  In all 3 cases the data collapsed onto what appears to be a single curve, particularly for small $q$ very close to the transition point.  For larger static fields of 200 mG, the difference between transverse and longitudinal instability rates was less appreciable, for reasons that are not presently clear.  It is possible that the larger microwave power required to cancel the increased quadratic shift caused a spin-state dependent atom loss that suppressed the instability.

The inset to the figure shows an expanded view of the data outlined in the dashed box.  Here, the data sets from different static fields have been combined into one.  This data very close to the transition point was separately fit to a Bogoliubov function $\Gamma(q-q0)$ that has been shifted empirically by $q0$, yielding $q0 = 0.65 \pm 0.13$ Hz.  The quoted error is the statistical uncertainty in the fit to the first 9 data points starting from the left in the inset.  Although no physical theory motivates this choice of fitting function, it does provide a useful parameterization of our data, particularly the {\em steepness} with which it approaches $\Gamma = 0$ from the $q<0$ side.  The error bars along the $q$-axis are the $200$ milliHertz experimental uncertainty in determination of the $q = 0$ point due to the bias magnetic field calibration uncertainty of 4 mG.  All other error sources, including microwave magnetic field amplitude and frequency uncertainties, were much smaller and could be neglected.  A better magnetic  field calibration could help reduce this uncertainty.  With this level of precision the data suggest a 2 to 3 $\sigma$ shift of the phase transition point to $q = +650$ milliHertz.  Some portion of this shift may be attributed to the finite, but small instability rate of $3$ s$^{-1}$ observed for $q>0$ in the absence of any microwave fields, where $q = + 3$ Hz.  This could be caused by background spin redistribution due to the small field gradients remaining after cancellation, as noted earlier, in conjunction with the finite thermal cloud.  Thermally induced distillation of atoms from $m_F = 0$ to $m_F = \pm 1$ has been previously observed in elongated BECs \cite{mies99meta}.  Notwithstanding these experimental caveats, the sub-Hz level precision spectroscopy of a phase transition has not been achieved previously in ultracold gases, to our knowledge.

While the data in Figure \ref{fig:Instability_Rate} appear to have a universal character, we do not yet have a complete explanation why transverse fields appear to suppress the instability relative to longitudinal fields. We have evidence, however, that spatial inhomogeneities in the field magnitude play an important role, and we explore this effect in the rest of the paper.  Since these gradients were different for transverse and longitudinal fields, this effect by itself might explain the observed differences.  

To understand this point in further detail, we note that for a one-dimensional system we only need to consider variations in magnetic field along the $y$ direction.  Thus for fields that are mostly $B_y$, the first order magnetic field is given by $B = |\vec{B}| \approx B_0 + \frac{\partial B_y}{\partial y} y$, and by applying an external field gradient $-\frac{\partial B_y}{\partial y}$ we could cancel the field inhomogeneity to first order.  For a bias field that is mostly $B_z$, the only term that is relevant in the same order is the variation of that field along $y$:  $|\vec{B}| \approx  B_0 + \frac{\partial B_z}{\partial y} y $, since all other gradient terms $\partial B_{y,x}/\partial y$ add in quadrature and should be suppressed.  A similar argument applies to fields that are mostly $B_x$.  Transverse field gradients of this type could neither be easily characterized nor cancelled using our current setup \footnote{This requires the installation of coils at an angle with respect to the principal $\hat{x},\hat{y},\hat{z}$ directions of our apparatus.}.  However, as noted earlier \cite{bookjans2011}, they did appear to play some role in the problem, since at long times $t > 1$ second, the cloud had separated into two distinct domains of $m_F = \pm 1$, consistent with a gradient in the linear Zeeman term.  In the absence of a field gradient these two spin states would be miscible with one another.

An alternative and very intriguing explanation is a genuine orientation dependence of the instability upon the bias field.  This would signal physics beyond a mean-field description of the spinor BEC, an exciting development.  For example, dipolar interactions \cite{pasquiou2011,eto2014,zhang2015} have an anisotropy in space and can influence the spin relaxation rate for sufficiently anisotropic trapping potentials \cite{deuretzbacher2010}.  The similarity of the data in Figure \ref{fig:Instability_Rate} for both $B_x$ and $B_z$ fields suggests this as a possibility, although the effect in Reference \cite{deuretzbacher2010} is unfortunately too weak to explain the factor of 10 suppression observed.  Without further study we hold dipolar effects in abeyance.

\begin{figure} [htbp]
\includegraphics[width = \columnwidth]{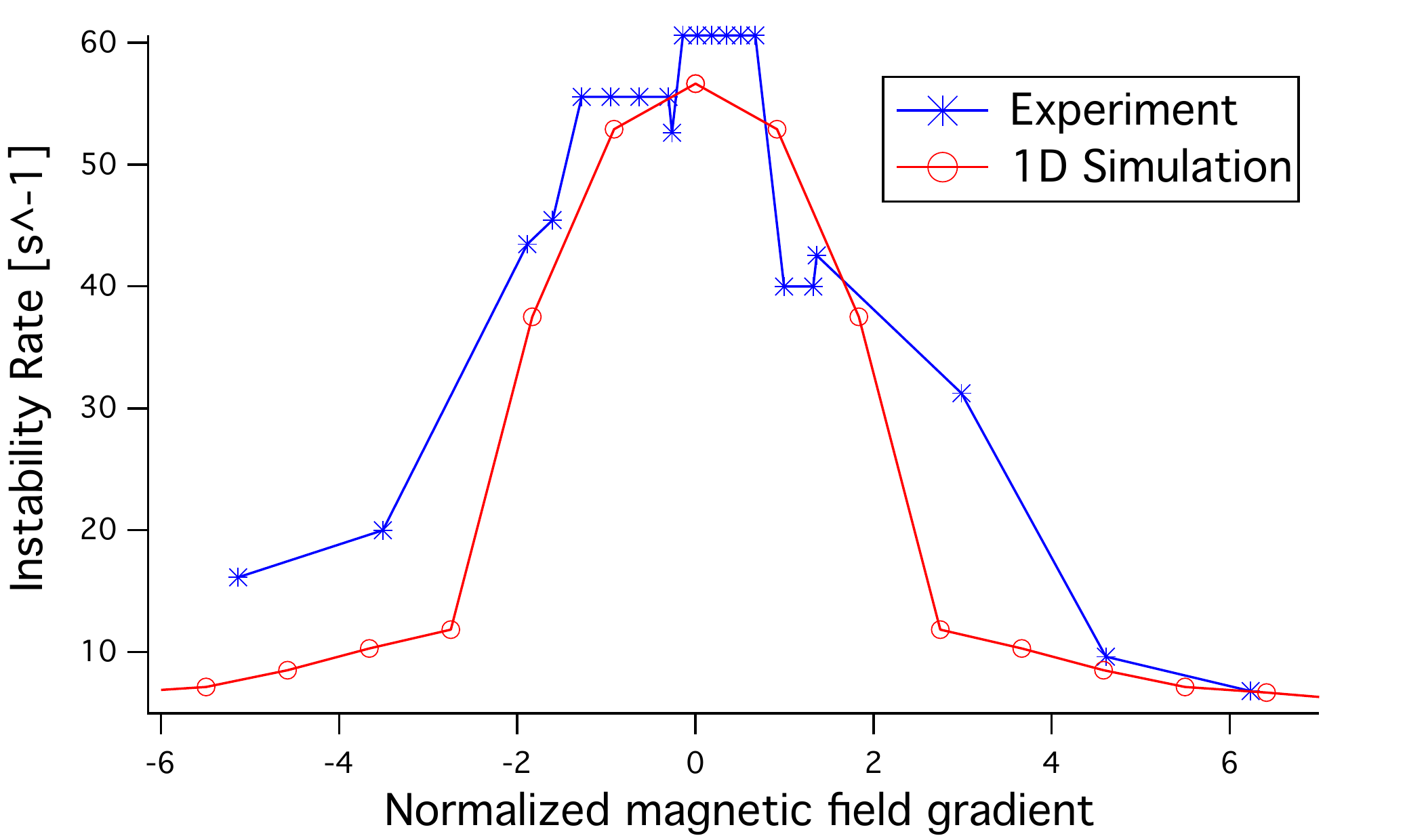}

\caption{(Color Online).  Field gradient slows down the instability. Shown are experimentally measured and numerically computed instability rates versus applied magnetic field gradient. Quadratic shift was $q = -3.7$ Hz at zero field gradient, and varied from -5 to -2.5 Hz over the data range shown.}
\label{fig:data_2}
\end{figure}

To separate out the role played by magnetic field gradients from other potential causes, we performed controlled experiments with the magnetic field applied along the $y$-direction, and negligible $z$ and $x$ fields.  Here the field gradient $B_p = \frac{\partial B_y}{\partial y}$ was {\em deliberately} applied, and could be tuned to both positive and negative values by varying the current in the anti-Helmholtz coils.  Thus, to a good approximation we had independent control over $p$ and $q$.  

Figure \ref{fig:data_2} shows the variation of $\Gamma_{1/2}$ with applied magnetic field gradient, which has been normalized in order to compare with 1D numerical simulations.  The normalized gradient is $B_p/B_p^0$, where $ B_p^0 = \frac{\hbar \omega_y}{g_F \mu_B a_y} \simeq 13$ mG/cm, where $\omega_y,a_y = \sqrt{\hbar/m\omega_y}$ are the axial frequency and oscillator length, respectively.  The data clearly show that the maximum rate occurs near $B_p = 0$, and falls off rapidly with field gradient to either positive or to negative values.  Due to non-idealities in the experiment, the gradient coils introduced an asymmetric bias field variation which was independently measured.  For normalized field gradients $<-3$ this caused an increase in $|q|$ that created a small, positive deviation between the experimental data and the theory on the left side of the graph.

Also plotted is the result of 1d numerical simulations based on the Truncated Wigner Approximation (TWA) \cite{vinit2013}.  These were performed for $B_p>0$ and the results reflected about the $y$-axis in the figure for $B_p<0$.  These numerical data were scaled by a factor of 2 in both $x$ and $y$ axis, and show good agreement with our measured data.  Although we cannot at present account for an overall scaling factor, we can account for the fact that it is the same for both x and y axes in the figure.  This is due to the linearity of the Bogoliubov equations that describe the initial instability.  All quantities of interest, including $p,q$ and the (imaginary) eigenvalues $E$, scale linearly with the chemical potential.  If experiment and theory were performed at different values for $\mu$, a single scaling factor should apply to the quantities plotted in both axes of Figure \ref{fig:data_2}.  This argument should be approximately true even for larger hold times, provided that the system is still in the growth phase of the dynamics where nonlinearities are not too strong. 
\begin{figure} [htbp]
\includegraphics[width = \columnwidth]{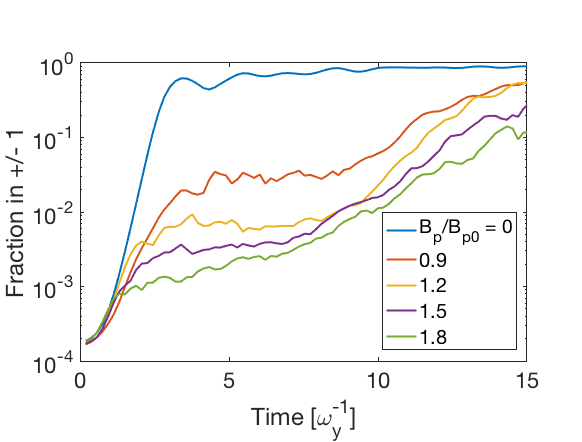}

\caption{(Color Online).  Origin of field-gradient induced suppression of instability.  Numerical simulations of the dynamical evolution of the $f_{\pm1}$ populations for various field gradients show the formation of a plateau, as described in the text.}
\label{fig:TWA}
\end{figure}

What causes the suppression?  The numerically obtained wavefunctions show that noise in the initial state becomes amplified by the instability, forming localized domains that grow with time, as noted in earlier work \cite{vinit2013}.  Figure \ref{fig:TWA} shows numerical results for the temporal dynamics of the population $f_{\pm 1}$ for different values of the field gradient $B_p$.  For both $B_p = 0$ and $B_p \neq 0$, we observe rapid domain growth, but for $B_p \neq 0$, a plateau in $f_{\pm 1}$ is reached.  The plateau value decreases with increasing field gradient, and is $< 0.01$ for normalized field gradients $B_p/B_p^0 > 3.3$.  Further growth of the $\pm 1$ populations must wait until a much longer time $T_{late} \sim 10 \omega_y^{-1}$, which is the timescale observed in the experiment, i.e., $\Gamma_{1/2} > 1/T_{late}$.  By examining the numerically generated wavefunctions, we observed that near $t = T_{late}$, the meagerly populated $\pm 1$ domains had diffused to opposite sides of the trap where their population could increase more easily at the expense of the smaller $m_F = 0$ population near the Thomas-Fermi boundaries.  

Our simulations therefore suggest that there are {\em two} stages to the dynamics--early ($t\ll T_{late}$) and late ($t \sim T_{late}$).  In the early stage, a clamping of the initial $m_F = \pm 1$ population, rather than a reduction of the instability rate, occurs.  This early stage is critical for slowing down the instability.  Unfortunately, our current experimental sensitivity does not allow us to probe population fractions $f_{\pm 1} < 0.05$.  
We can, however, understand the numerical observations in terms of a decoherence process caused by the field gradient.  In the presence of a magnetic field gradient, the quantum field operator, $\hat{\psi}_m$, acquires a spatially varying phase $\phi = m E_Z t y/\hbar$, where $E_Z = g_F \mu_B (dB/dy)$.  If $\phi$ varies by $2 \pi$ over a single domain, the effective rate of amplification can be reduced by destructive interference from different spatial regions.  For a domain of size $d$ this occurs when $\frac{B_p}{B_{p0}} \left (\frac{\omega_y t d}{a_y} \right ) = \pi$.  For $\omega_y = 2 \pi \times 7 \rm{s}^{-1}$, a domain size of $30 \mu$m and $\frac{B_p}{B_{p0}} = 1.2$, this occurs at $16$ ms, comparable with the timescale of the instability $T_{1/2}$ for $B_p = 0$ (see Figure \ref{fig:data_2}).  This picture is therefore consistent with the formation of a plateau early in the dynamical evolution, as seen in Figure \ref{fig:TWA}.  Since the decoherence is a process local to individual domains, it should not depend on whether the overall density profile is homogeneous or inhomogeneous.

In conclusion, we have made sub-Hz level precise measurements of the location of a quantum phase transition by observing a dynamical instability.  We achieved this through careful control of magnetic field gradients that revealed a new mechanism for suppression of the instability.  With the gradient removed, the extreme sensitivity of this phase transition to quadratic shifts could be a new tool for performing precise magnetometry with spinor Bose-Einstein condensates.  It is immune to density fluctuations, in contrast to schemes relying on imaging of condensate motion \cite{yang2016}, and has potentially different quantum noise limits than Larmor precession \cite{vengalattore07}.

We thank Mukund Vengalattore and Carlos S\'a de Melo for useful conversations.  This work was supported by NSF grant No.\ 1100179.

\bibliographystyle{apsrev}

\end{document}